\documentclass[aps,prl,twocolumn,superscriptaddress]{revtex4}

\usepackage{graphicx}
\usepackage{textcomp}
\usepackage{amsmath}
\usepackage{bm}
\usepackage[small]{subfigure}
\graphicspath{{figures/}}

\begin{document}

\title{Enhanced solid-state multi-spin metrology using dynamical decoupling}

\author{L. M. Pham}
  \affiliation{School of Engineering and Applied Sciences, Harvard University, Cambridge, MA 02138, USA}
\author{N. Bar-Gill}
  \affiliation{Harvard-Smithsonian Center for Astrophysics, Cambridge, MA 02138, USA}
  \affiliation{Physics Department, Harvard University, Cambridge, MA 02138, USA}
\author{C. Belthangady}
\author{D. Le Sage}
  \affiliation{Harvard-Smithsonian Center for Astrophysics, Cambridge, MA 02138, USA}
\author{P. Cappellaro}
  \affiliation{Nuclear Science and Engineering Department, Massachusetts Institute of Technology, Cambridge, MA 02139, USA}
\author{M. D. Lukin}
  \affiliation{Physics Department, Harvard University, Cambridge, MA 02138, USA}
\author{A. Yacoby}
  \affiliation{Physics Department, Harvard University, Cambridge, MA 02138, USA}
\author{R. L. Walsworth}
  \affiliation{Harvard-Smithsonian Center for Astrophysics, Cambridge, MA 02138, USA}
  \affiliation{Physics Department, Harvard University, Cambridge, MA 02138, USA}

\date{\today}

\begin{abstract}
We use multi-pulse dynamical decoupling to increase the coherence lifetime ($T_2$) of large numbers of nitrogen-vacancy (NV) electronic spins in room temperature diamond, thus enabling scalable applications of multi-spin quantum information processing and metrology. We realize an order-of-magnitude extension of the NV multi-spin $T_2$ for diamond samples with widely differing spin environments. For samples with nitrogen impurity concentration $\lesssim 1$ ppm, we find $T_2$ $>2$ ms, comparable to the longest coherence time reported for single NV centers, and demonstrate a ten-fold enhancement in NV multi-spin sensing of AC magnetic fields.
\end{abstract}


\maketitle

The negatively-charged nitrogen-vacancy (NV) color center in diamond possesses many useful properties---long electronic spin coherence times at room temperature, an optical mechanism for initializing and detecting the spin state, and the ability to coherently manipulate the spin using electron spin resonance (ESR) techniques---which make it a leading solid-state system for scalable applications in quantum information and metrology, such as sensitive detection of electric and magnetic fields with high spatial resolution~\citep{TaylorNatPhys2008, MazeNat2008, BalasubramanianNat2008, DegenAPL2008, MaurerNatPhys2010, GrinoldsNatPhys2011, MaletinskyArXiv2011, DoldeNatPhys2011} or in bulk~\cite{AcostaAPL2010,Shin2012}. Recently, dynamical decoupling techniques have been used to reduce the effective interaction of single NV spins with other spin impurities in the environment, enabling significant improvements in the NV single-spin coherence lifetime ($T_2$) ~\citep{deLangeSci2010, RyanPRL2010, NaydenovPRB2011} and AC magnetic field sensitivity~\citep{deLangePRL2011, NaydenovPRB2011}. In this Letter, we demonstrate the successful application of dynamical decoupling to large numbers of NV spins ($>10^3$) in room temperature diamond~\citep{TaylorNatPhys2008, StanwixPRB2010, PhamNJP2010}. We employ multi-pulse Carr-Purcell-Meiboom-Gill (CPMG) and XY control sequences~\citep{CarrPR1954,MeiboomRSI1958} to improve both the NV multi-spin $T_2$ and AC magnetic field sensitivity by an order of magnitude. We find similar relative improvements for diamond samples with widely differing NV densities and spin impurity concentrations.  For some samples, the NV multi-spin $T_2$ is increased to $>2$ ms, where it begins to be limited by NV spin-lattice relaxation ($T_1 \approx 6$ ms). These demonstrations of the utility of dynamical decoupling for solid-state multi-spin systems pave the way for scalable applications of quantum information processing and metrology in a wide-range of architectures, including multiple NV centers in bulk diamond~\citep{TaylorNatPhys2008, StanwixPRB2010}, 2D (thin-layer) arrays~\citep{MaurerNatPhys2010,PhamNJP2010,SteinertRevSciIns2010,MaertzAPL2010}, and diamond nano-structures~\citep{BabinecNatNano2010, BaranovSmall2011}; as well as phosphorous donors in silicon~\citep{TyryshkinPRB2003, SimmonsNature2011} and quantum dots~\citep{HansonRMP2007}.

\begin{figure}[]
\includegraphics[width=0.91\columnwidth]{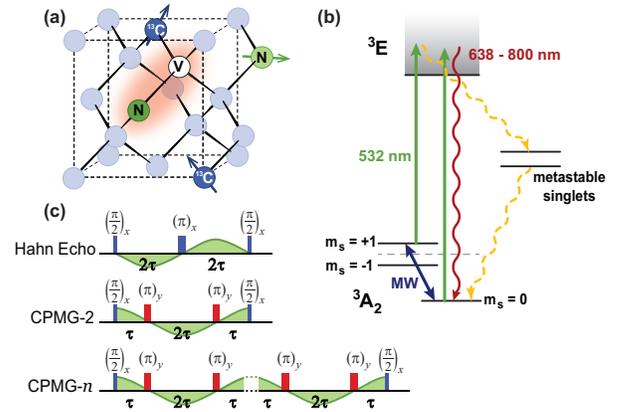}
\label{fig:NV}
\caption{(a) Nitrogen vacancy (NV) color center in a diamond lattice. NV electronic spin decoherence is dominated by $^{13}$C nuclear spin and N electronic spin impurities. (b) Energy level structure of negatively charged NV center. (c) Hahn Echo and $n$-pulse CPMG control sequences. Timing of AC magnetic field to be measured is shown in green.}
\end{figure}

The NV center is composed of a substitutional nitrogen (N) impurity and a vacancy (V) on adjacent lattice sites in the diamond crystal (Fig. 1(a)). The electronic structure of the negatively charged state of the NV center has a spin-triplet ground state, where the $m_s = \pm 1$ levels are shifted from the $m_s = 0$ level by $\sim 2.87$ GHz due to the dipolar spin-spin interaction (Fig. 1(b)). Application of an external static magnetic field along the N-V axis Zeeman shifts the $m_s = \pm 1$ levels and allows one to treat the $m_s = 0$, $m_s = +1$ spin manifold (for example) as an effective two-level system. The NV spin state can be initialized in the $m_s = 0$ state with above-band laser excitation, manipulated with resonant microwave (MW) pulses, and read out optically by measuring the spin-state-dependent fluorescence intensity in the phonon sideband.

The NV spin environment (i.e., spin bath) is dominated by $^{13}$C and N impurities, randomly distributed in the diamond crystal. These spin impurities create time-varying local magnetic fields at each NV spin, which can be approximated as an average local magnetic field that fluctuates on a timescale set by the mean interaction between spins in the bath, inducing rapid dephasing of freely precessing NV spins on a timescale $T_2^* \sim 1\ \rm{\mu s}$ for typical spin impurity concentrations.  By applying a single resonant microwave $\pi$ pulse to refocus the dephasing, the Hahn Echo sequence (Fig. 1(c)) decouples NV spins from bath field fluctuations that are slow compared to the free precession time~\citep{TaylorNatPhys2008,MazeNat2008}. Application of additional control pulses, as in $n$-pulse CPMG (CPMG-$n$) and XY sequences, have recently been shown to decouple single NV spins from higher frequency bath fluctuations~\citep{deLangeSci2010, RyanPRL2010, NaydenovPRB2011}.

In the present study, we used a wide-field fluorescence microscope to perform dynamical decoupling and magnetometry measurements on large numbers of NV centers in three diamond samples with different NV densities and spin impurity environments. A switched 3-Watt 532-nm laser provided optical excitation of NV centers within a $\rm{10\ \mu m}$-diameter cross-section of each sample. NV spin-state-dependent fluorescence was collected by a microscope objective and imaged onto a CCD array. Resonant microwave control pulses for coherent manipulation of the NV spin states were applied using a loop antenna designed to generate a homogeneous $\rm{B_1}$ field over the sample detection volume. Applying a static field ($\rm{B_0 \sim 70}$ Gauss) along one of the four diamond crystallographic axes selected approximately one quarter of the NV centers to be resonant with the microwave pulses. Each diamond sample consisted of an NV-rich layer grown by chemical vapor deposition on a non-fluorescing diamond substrate, such that all collected fluorescence could be attributed to the NV-rich layer.

\begin{figure}[!ht]
\includegraphics[width=0.91\columnwidth]{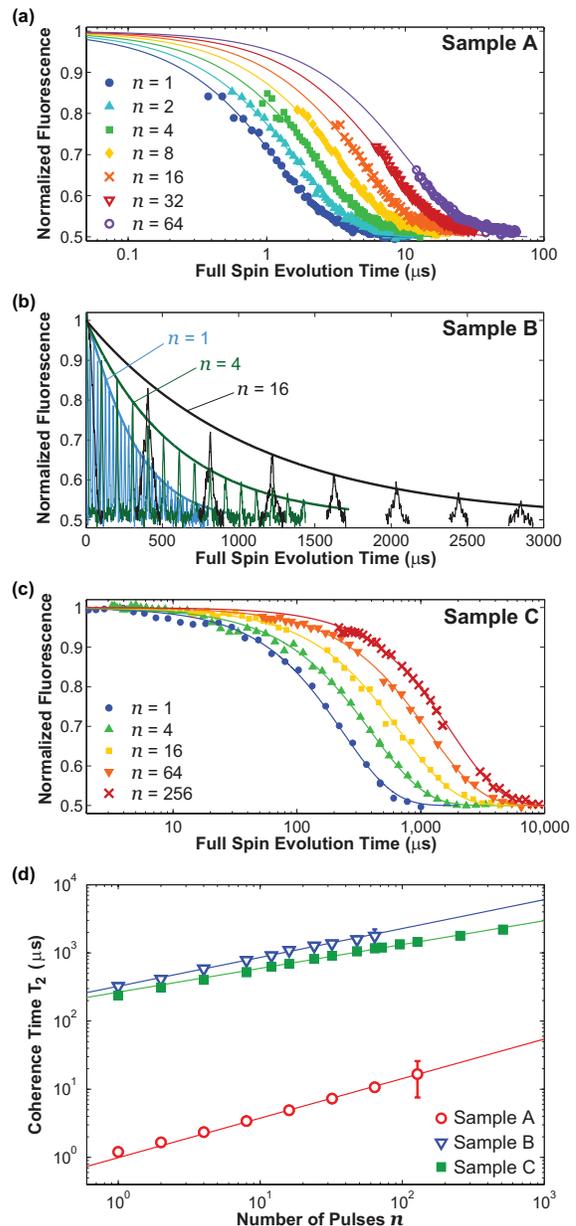}
\label{fig:T2DD}
\caption{Measurements of NV multi-spin coherent evolution using $n$-pulse CPMG control sequences for three diamond samples of differing NV densities and spin impurity environments: (a) NV $\sim 60$ ppb, N $\sim 100$ ppm, and 1.1\% $\rm{^{13}C}$; (b) NV $\sim 0.2$ ppb, N $\sim 0.1$ ppm, and 1.1\% $\rm{^{13}C}$; and (c) NV $\sim 0.6$ ppb, N $\sim 1$ ppm, and 0.01\% $\rm{^{13}C}$, where the solid lines denote fits to the decoherence envelope of the form $\exp\left[-\left(\tau/_{T_2}\right)^p\right]$. Note that the time axis of (b) is plotted with a linear scale due to the periodic collapses and revivals in the NV spin coherence of Sample B associated with $\rm{^{13}C}$ Larmor precession~\citep{ChildressSci2006}. These collapses and revivals occur in samples where the $^{13}$C abundance is high enough to contribute significantly to NV decoherence. (d) Scaling of measured NV multi-spin coherence times with the number $n$ of CPMG pulses: $T_2 \propto (n)^{s}$.}
\end{figure}

Sample A (Apollo) had a 16 $\rm{\mu m}$-thick NV-rich layer with NV concentration $\sim 60$ ppb (measured by NV fluorescence intensity), N concentration $\sim 100$ ppm (measured by secondary ion mass spectroscopy), and 1.1\% natural abundance $\rm{^{13}C}$ concentration. The high N concentration dominated NV decoherence in this sample, limiting the measured Hahn Echo multi-spin coherence time to $T_2 \rm{\approx 2\ \mu s}$. We applied CPMG-$n$ sequences and determined the NV multi-spin coherence time as a function of the number of pulses, $T_2^{(n)}$, from the $1/e$ decay  of the spins' coherent evolution as a function of the total CPMG-$n$ evolution period (Fig. 1(c)). Representative measurements of NV multi-spin coherence decay are shown in Fig. 2(a), with $T_2^{(n)}$ extended by a factor $>10$ for $n = 128$ (Fig. 2(d)). Furthermore, we found that $T_2^{(n)}$ exhibited a power-law dependence on $n$: $T_2^{(n)} \propto n^s$, with $s = 0.65 \pm 0.02$  for Sample A, which is consistent with the value $s \approx 0.67$ found recently for single NV centers in similarly nitrogen-rich diamond samples~\citep{deLangeSci2010}. These results demonstrate that inhomogeneities in the spin bath do not limit the effectiveness of dynamical decoupling for extending solid-state multi-spin coherence times by at least an order of magnitude.

The NV-rich layers of Samples B (Apollo) and C (Element Six) had much lower NV and N concentrations than in Sample A: NV $\sim 0.2$ ppb, N $\sim 0.1$ ppm for Sample B and NV $\sim 0.6$ ppb, N $\sim 1$ ppm for Sample C. However, while Sample B contained 1.1\% natural abundance $^{13}$C, Sample C was isotopically engineered to reduce the $^{13}$C concentration to 0.01\%. Despite these differences in spin impurity environments, we measured similar Hahn Echo NV multi-spin coherence times ($\rm{T_2 \sim 300\ \mu s}$) for the two samples. By applying $n$-pulse CPMG dynamical decoupling sequences, we extended the multi-spin $T_2^{(n)}$ to $\rm{\approx 2}$ ms for both samples (Figs. 2(b) and 2(c)), which is comparable to the longest coherence time reported for dynamical decoupling applied to single NV centers~\citep{NaydenovPRB2011} and beginning to be limited by NV spin-lattice relaxation ($T_1 \approx 6$ ms).

For Samples B and C we also observed a power-law dependence for $T_2^{(n)} \propto n^s$ (Fig. 2(d)), with lower scaling powers than for the N-rich Sample A: $\rm{s = 0.42 \pm 0.02}$ for Sample B and $\rm{s = 0.35 \pm 0.01}$ for Sample C. These sample-dependent scaling powers suggest that multi-pulse dynamical decoupling control sequences can serve as spectral decomposition probes of spin bath dynamics~\citep{BarGill2011}.

We next applied dynamical decoupling to improve the sensitivity of NV multi-spin magnetometry. In a standard AC magnetometry measurement utilizing a Hahn Echo sequence, an oscillating magnetic field, $b(t) = b_{ac} \sin{\left[ \left(2 \pi/\tau_{ac}\right) t + \phi \right]}$, induces a net phase accumulation of the NV spin coherence, which is maximized when the full time of the Hahn Echo sequence is equivalent to the period of the AC magnetic field ($\tau_{ac}$) and the phase offset $\phi$ is such that the control pulses coincide with nodes in the magnetic field (Fig. 1(c)). Under these conditions, the field amplitude $b_{ac}$ can be extracted from the measurement of accumulated NV spin phase with optimum sensitivity, given approximately by~\citep{TaylorNatPhys2008}:
\begin{equation}\label{eq:HEsens}
\eta_{HE} \approx \frac{\pi \hbar}{2 g \mu_B} \frac{1}{C \sqrt{\tau_{ac}}} \exp{\left[\left(\frac{\tau_{ac}}{T_2}\right)^p\right]}.
\end{equation}
Here $C$ is a parameter that encompasses the measurement contrast, optical collection efficiency, and number of NV spins contributing to the measurement. The contrast is modified by NV decoherence over the course of the measurement, described phenomenologically by an exponential factor with power $p$. The value of $p$ is found to be sample dependent, in the range of 1 to 2.5, and is related to the dynamics of the spin environment and to ensemble inhomogeneous broadening~\citep{BarGill2011}.

In an AC magnetometry measurement utilizing $n$-pulse dynamical decoupling, the sensitivity is given approximately by Eq.~\ref{eq:HEsens} with two modifications: (1) the measurement time is increased by $\tau_{ac} \rightarrow \frac{n}{2}\tau_{ac}$, and (2) the NV multi-spin coherence time is extended by $T_2 \rightarrow T_2 n^s$. The resulting sensitivity is given by:
\begin{equation}\label{eq:MPsens}
\eta_{(n)} \approx \frac{\pi \hbar}{2 g \mu_B} \frac{1}{C \sqrt{\frac{n}{2}\tau_{ac}}} \exp{\left[\left(\frac{n^{(1-s)}\tau_{ac}}{2 T_2}\right)^p\right]}.
\end{equation}

Because the measurement time increases linearly with the number of control pulses $n$, whereas the coherence time increases sub-linearly, there is an optimum number of pulses $n_{opt}$ that yields the most sensitive measurement of an AC magnetic field of period $\tau_{ac}$ given a set of sample-determined parameters:
\begin{equation}\label{eq:Nopt}
n_{opt} = \left[\frac{1}{2 p (1-s)} \left(\frac{2 T_2}{\tau_{ac}}\right)^p\right]^{\frac{1}{p(1-s)}}.
\end{equation}

For a given sample, all the parameters except $\tau_{ac}$ are fixed and we can simplify Eq.~\ref{eq:Nopt} to $n_{opt} \propto \left(1/\tau_{ac}\right)^{\frac{1}{(1-s)}}$. From this relationship, we see that at higher AC frequencies, more pulses are needed to reach the optimum sensitivity, which can be understood intuitively by realizing that the high frequency regime corresponds to short time intervals between control pulses during which time there is very little contrast lost due to decoherence ($\tau_{ac} \ll T_2$). More pulses increase the sensitivity by allowing for a longer measurement time and subsequently more phase accumulation per measurement. This intuition also illustrates why multi-pulse sequences are more effective at enhancing magnetometry sensitivity in the high-frequency regime, where the Hahn Echo scheme provides relatively poor magnetic field sensitivity (Fig. 3(c)). Note that extension of the NV multi-spin coherence time via multi-pulse dynamical decoupling (and thus enhancement of magnetic field sensitivity) is eventually limited by NV spin-lattice relaxation ($T_1$), beyond which increasing the number of control pulses is ineffective.

\begin{figure}[!ht]
\includegraphics[width=0.91\columnwidth]{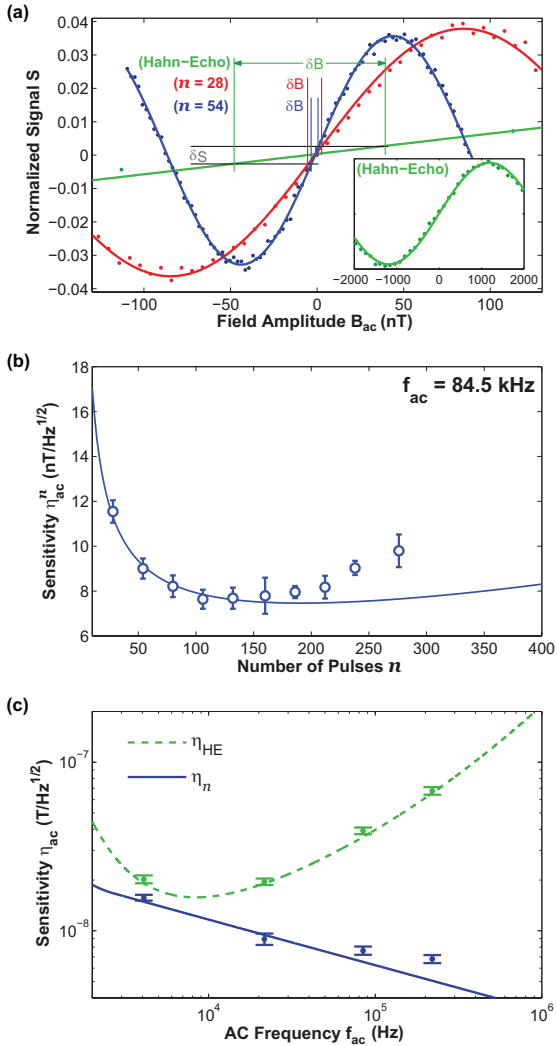}
\label{fig:Mgtry}
\caption{Measured AC magnetic field sensitivity for a 30 $\mu m^3$ sensing volume of Sample C ($\sim 10^3$ sensing NV spins). (a) Examples of measured normalized fluorescence signals as functions of AC field magnitude $b_{ac}$ using a Hahn Echo sequence (green) (wider amplitude range shown in the inset) and multi-pulse XY sequences with 28 (red) and 54 (blue) control pulses, illustrating how the uncertainty in the measured signal ($\delta S$) limits the uncertainty in the extracted magnetic field magnitude ($\delta B$). The sine behavior of the signal with respect to $b_{ac}$ is achieved by shifting the phase of the last microwave pulse by $90^{\circ}$ from what is shown in Fig. 1(c). (b) Sensitivity as a function of number of pulses for an 84.5 kHz AC field (circles). The solid line is calculated using Eq.~\ref{eq:MPsens} with experimental parameters $C = 0.1$, $T_2 = 250\ \mu s$, $p = 1$, and $s = 0.37$. (c) Comparison of calculated (lines) and measured (points) sensitivity at several AC magnetic field frequencies.}
\end{figure}

We used Sample C to perform NV multi-spin magnetometry measurements comparing Hahn Echo and multi-pulse dynamical decoupling schemes. The diamond detection volume was approximately $30\ \mu m^3$, corresponding to $\sim 10^3$ sensing NV centers aligned along the static magnetic field. We employed $n$-pulse XY sequences, in which control pulses are applied with the same timing as in CPMG-$n$ sequences but with alternating $90^{\circ}$ spin rotation axes to provide more isotropic compensation for pulse errors~\citep{GullionJMR1990}. Fig. 3(a) illustrates how we extracted the magnetic field sensitivity from the measured variation of NV fluorescence as a function of applied AC magnetic field amplitude~\citep{TaylorNatPhys2008, MazeNat2008} for different measurement pulse sequences. The AC magnetic field sensitivity improved with the number of pulses, in good agreement with predicted values up to $n \approx 150$ pulses, at which point pulse infidelities began to degrade the measured sensitivity (Fig. 3(b)).

Over a wide range of AC magnetic field frequencies our NV multi-spin measurements confirmed that multi-pulse dynamical decoupling outperformed the Hahn Echo scheme, in agreement with theoretical expectations (Fig. 3(c)). The enhancement in magnetic field sensitivity provided by multi-pulse dynamical decoupling was especially pronounced at frequencies higher than the Hahn Echo $1/T_2$ coherence. For example, at a frequency of 220 kHz, we measured a factor of 10 improvement in magnetic field sensitivity: from $\rm{67.7 \pm 3.5 nT/\sqrt{Hz}}$ using a Hahn Echo sequence, to $\rm{6.8 \pm 0.4 nT/\sqrt{Hz}}$ using a 240-pulse XY sequence.

In summary, we experimentally demonstrated that multi-pulse dynamical decoupling sequences extend by an order of magnitude the coherence lifetime of large numbers ($>10^3$) of NV electronic spins in room temperature diamond, for samples with widely-differing NV densities and spin environments. We realized an extension of the NV multi-spin coherence time to $>2$ ms, which is comparable to the best results from application of dynamical decoupling to single NV centers. We also showed that multi-pulse dynamical decoupling improves NV multi-spin AC magnetic field sensitivity relative to the Hahn Echo scheme, with a ten-fold enhancement for higher frequency fields. Further improvements in NV multi-spin coherence time and magnetic field sensitivity are expected from the integration of dynamical decoupling with diamond samples engineered to have optimized spin environments, as well as quantum-assisted techniques~\citep{JiangScience2009} and increased optical collection efficiency~\citep{BabinecNatNano2010,LeSage2012}. The present results aid the development of scalable applications of quantum information and metrology in a variety of solid-state multi-spin systems, including NV centers in bulk diamond, 2D arrays, and nanostructures; phosphorous donors in silicon; and quantum dots.

\begin{acknowledgments}
This work was supported by NIST, NSF and DARPA (QuEST and QuASAR programs). We gratefully acknowledge the provision of diamond samples by Element Six and Apollo and helpful technical discussions with Daniel Twitchen and Matthew Markham.
\end{acknowledgments}

\bibliography{lmpham}

\end{document}